# Mobile Agent Systems, Recent Security Threats and Counter Measures

Belal Amro

Computer Science Department, College of Science and Technology, Hebron University
Hebron, West Bank , Palestine
*bilala@hebron.edu*

**Abstract**

Mobile agent technology offers a dramatically evolving computing paradigm in which a program, in the form of a software agent, can suspend its execution on a host computer, transfers itself to another agent-enabled host on the network, and resumes execution on the new host. It is 1960's since mobile code has been used in the form of remote job entry systems. Today's mobile agents can be characterized in a number of ways ranging from simple distributed objects to highly organized intelligent softwares. As a result of this rapid evolvement of mobile agents, plenty of critical security issues has risen and plenty of work is being done to address these problems. The aim is to provide trusted mobile agent systems that can be easily deployed and widely adopted. In this paper, we provide an overview of the most recent threats facing the designers of agent platforms and the developers of agent-based applications. The paper also identifies security objectives, and measures for countering the identified threats and fulfilling those security objectives.

## 1. Introduction

Software agents are a very promising evolving computing paradigm. However, there is no precise definition of a mobile agent till now. Some popular definitions of mobile agents are: "a persistent software entity dedicated for a special purpose" [1], " a software entity which tasks can be delegated" [2], and " computer programs that simulate a human relationship by doing something that another person could do for you" [3]. A comprehensive definition can be summarized as "Mobile Agent is a program that can exercise an individual's or organization's authority, work autonomously toward a goal, and meet and interact with other agents".[4].
A mobile agent consists of three components namely code, state, and attributes. The code is the set of instructions that defines the agent's behavior. The state describes the agent's internal variables and enables it to resume its activities after moving to another host. Finally, attributes are information describing the agent, its owner, its movements, resource, and keys[ 5].
Mobile agents may interact among each other via a contract and service negotiation, auctioning, and bartering. They may also be either stationary, always resident at a single platform; or mobile, capable of moving among different platforms at different times. [6]

The rest of the paper is organized as follows: Section 2 describes the characteristics of mobile agents. In section 3, we will discuss recent security threats on mobile agents. Section 4 will cover the countermeasures and security enhancements techniques for security threats on mobile agents. Conclusions and Future work will be summarized in Section 5.

## 2. Mobile agent characteristics and system models

Due to rapid growth and great advantages mobile agents bring to computing, plenty of work were done either on the mobile agent itself or on mobile agent systems holistically. Here we will provide the basic characteristics of a mobile agent as well as mobile system models as well.

2.1 Mobile Agent Characteristics

As described in Section 1, there is no precise definition of a mobile agent. However, every definition brings a group of characteristics of a mobile agent. A summary of these characteristics is provided below, these characteristics are extracted from [6], [7], [8], and [9] definitions of mobile agents.

- **Autonomist:** the ability of the agent to execute without the need for human interaction. This feature does not prevent intermittent interaction which might be required from time to time.
- **Intelligence:** the ability of the agent to learn, and adapt over time. Learning is crucial for intelligent mobile systems and enables them to adapt their behavior accordingly.
- **Communicative:** an agent should have the ability to communicate with other agents for the purpose

of exchanging data. This communication should be regulated and monitored some how to prevent security breaches.
- **Goal Oriented:** The mobile agent should have been oriented to a achieve a goal. This goal is explicitly stated in its internal plan of action.
- **Mobility:** Mobile agent can decide to migrate to a different machine or network while maintaining their persistence (consistent internal state over time)
- **Perceiving:** A mobile agent should perceive its surrounding environment and react or response accordingly. Sometimes agents should not just react, they may take active steps to change that environment according to their own desire.

2.2 Advantages of Mobile Agent Characteristics:

Mobile agents are independent and can respond to changes in environment. This provides a strong basis to build up a reliable and robust system. Hence agents are used to ease user tasks and adapt to user requirements. However, this advantage of using mobile agents led to a very complex structure for mobile agent interactions. Besides, the dynamic nature of agents made it difficult to predict the agent behavior. This difficulty may lead to severe security problems that will be discussed in Section 3.

2.3 Agent System Models:

Agent systems can be categorized into four groups[10]:

- **Client Server Model (CS):** The client component *A* requests the execution of a service with an interaction with the server component *B*. As a response, *B* performs the requested service by executing the corresponding know-how and accessing the involved resources collocated with *B*.
- **Remote Evaluation (REV):** In REV paradigm, a component *A* has code to be executed but it lacks the resources required, which happen to be located at a remote site *B*. Consequently, *A* sends the code to *B* located at the remote site. *B*, in turn, executes the code using the resources available there and returns results to A.
- **Code On Demand (COD):** In the COD paradigm, component *A* is already able to access the resources it needs. However, no information about how to manipulate such resources is available at *A*. Thus, *A* interacts with a component *B* at *SB* by requesting the code to be executed.
- **Mobile Agent (MA):** The *mobile agent* paradigm is different from other mobile code paradigms since the associated interactions involve the mobility of an *existing* computational component. In other words, while in REV and COD the focus is on the transfer of code between components, in the mobile agent paradigm a whole computational component is moved to a remote site, along with its state, the code it needs, and some resources required to perform the task.

In this paper, we will discuss security issues related to mobile agents, the mobile agent model consisting of a mobile agent and an agent platform is shown in Figure 1.

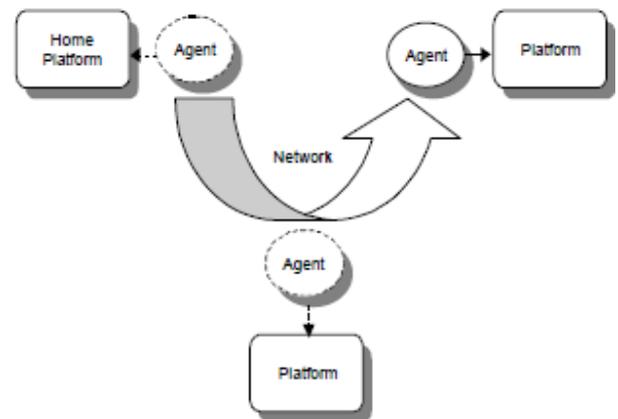

Fig. 1 Mobile agent model

## 3. Mobile Agent Security Threats

The dynamic nature of mobile agent led to complex design and security threats. These threats are categorized in four categories according to the originator of the threat and the victim. These categories are listed below:
1. **Agent to Platform:** This category is related to threats of a mobile agent to a particular platform.
2. **Platform to Agent**: This category is related to threats of platform to a particular mobile agent.
3. **Agent to Agent:** This category is related to threats that may occur due to the interaction among mobile agents.
4. **Platform to Platform:** This category addresses threats among platforms of mobile agents.

Researchers try to pin out each category and explain related threats. However, and because many threats span different categories, we will explain these threats independently and then show where they may occur. In Section 4, we will provide countermeasures against these threats.

### 3.1 Masquerading:

A masquerade attack occurs when an illegitimate user tries to impersonate a legitimate user; therefore, the masquerade user gets the privileges from the legitimate user account. In mobile agent systems, this class of attacks may occur in the four above listed threat categories.
In agent to platform and platform to agent threat categories, an unauthorized agent may claim the identity of another agent and hence gains access to services and resources to which it is not entitled. The same way, an agent platform can masquerade as another platform and hence deceives the mobile agent to its true destination. This will enable the fake platform extract sensitive information from deceived agents[11], [12].

Agents can communicate with each other and hence a malicious agent can deceive another agent and extract some sensitive information from it. The same can be done by platforms, a remote fake platform can deceive a legitimate one, and the latter may exchange information and forward agents to the faked platform[13], [14].

### 3.2 Unauthorized Access:

This type of attack is applicable for agent to agent and agent to platform, and platform to agent threat classes. By this attack a mobile agent can have access to a particular platform and hence affect other legitimate agents. Besides a mobile agent can directly interfere with another agent by invoking its public methods. It may even access and modify the agent's data or code. This alternation may affect and change the legitimate agent's behavior (e.g., turning a trusted agent into a malicious one). [3]

### 3.3 Denial of Service:

Commonly, the term denial of service is used for attacks in which the focus is on exhausting resources with the effect that other entities cannot be served anymore. This definition works well for agent platforms. However, in mobile agent systems, this attack definition is extended to preventing an agent from continuing to migrate to another host or may even delete the agent.[15]. this attack is applied to the platform to agent, agent to agent , and agent to platform threat classes.

### 3.4 Repudiation:

Repudiation attacks refer to denial of participation in the communication or transaction. Repudiation can lead to serious disputes that may not be easily resolved unless proper counter measures are applied. In Section 4, we will report these countermeasures for such types of attacks.

### 3.5 Alternation:

Platform to agent threat class is subject to this type of attack. This attack implies the malicious alternation of the code or data of a mobile agent without being detected. Detection of malicious alternation is not that simple and does not have a general solution till now because the visiting agent platform have the right to access some of the code and data of a mobile agent and consequently it may alter them.

### 3.6 Eavesdropping:

Eavesdropping is a passive attack which involves the interception and monitoring of secret communications of a mobile agent. In mobile agents, eavesdropping is more sever because a platform can monitor every instruction executed by the agent. This forms a strong heuristic about the behavior of that agent.

Table 1 below relates these attacks to threat classes.

Table 1: summary of threat classes and corresponding attacks

| Attack | Threat class | | | |
|---|---|---|---|---|
| | *1* | *2* | *3* | *4* |
| *Masquerading* | Yes | Yes | Yes | Yes |
| *Unauthorized* | Yes | Yes | Yes | No |
| *Denial of Service* | Yes | Yes | No | No |
| *Repudiation* | Yes | Yes | No | No |
| *Alternation* | No | Yes | Yes | No |
| *Eavesdropping* | No | Yes | No | No |

## 4. Mobile agent security countermeasures:

In this Section, we will discuss security countermeasures required for mobile agent systems. First, we will describe the basic security requirements, and then we will list some dedicated security mechanisms used to secure mobile agent systems and detect or prevent some attacks.

## 4.1 Basic Security Requirements For M.A. Systems:

In this subsection, we will provide the basic requirements for securing mobile agent systems, these include: authentication, confidentiality, availability, and accountability and non-repudiation.

### 4.1.1 Authentication and Authorization:

Authentication is the process of verifying the identity of the entity. In mobile agent systems, authentication process requires both agent and platform to be authenticated by each other. i.e. the agent knows the executing environment and the executing environment knows the agent. Authorization is the process of deciding to grant a request or not after entity has been authenticated. To achieve those security properties, digital signatures and password protection are used together.

### 4.1.2 Confidentiality, Privacy, and Anonymity:

Confidentiality refers to the state of hiding sensitive data from being disclosed to un-authorized parties. The disclosure of such data may degrade the privacy level since data may have private information about agent. Revealing the behavior of a mobile agent may also degrade privacy to some extent. The privacy concerns may be treated by means of privacy preserving techniques such as enforcing an anonymity level which might reduce the threat [16]. Encryption also works well for hiding sensitive data from un-authorized parties but it may degrade performance[17].

### 4.1.3 Accountability and Non-repudiation:

The problem of repudiation arise when a party claims being not involved in activity or a communication while it actually did. To overcome this problem, important communications and security related activities should be securely recorded for auditing and tracing purposes as well as for non-repudiation. These logs must be protected from un-authorized access to maintain the privacy and security levels of the system.

### 4.1.4 Availability

A mobile agent platform should ensure the availability of data and services required for local and incoming mobile agents. This implies that the platform should provide controlled concurrency, simultaneous access, deadlock management and exclusive access when required.[18].

Platforms should also be able to detect and recover from software crashes as well as hardware failures. It should also deal well and defend against denial of Service attacks.

## 4.2 Security Threat Detection and Prevention Mechanisms in Mobile Agent Systems:

In this sub-section, we will list some proposed mechanisms to detect or prevent an attack on mobile agent systems. These mechanisms are used to ensure that platforms will respect the policies of mobile agents they serve.

### 4.2.1 Detection Techniques:

Detection techniques are used to find out whether an agent has been changed or not. This includes tampering code, state, or execution flow. These techniques varies according to whether they work automatically or not, whether they work during execution or after termination, and whether they detect all possible alternations or some of them[19], [20].

Detection mechanisms also varies according to the scope of detection, some techniques use range checkers which detects illicit code manipulation according to variable values or timing constraints. Others use execution tracing and cryptography that allows them to detect attacks against code, state, and execution flow of mobile software components [21]. Hash function has been proposed as a detection mechanism for protecting the forward integrity results gathered by mobile programs [22], [23].

### 4.2.1 Prevention Techniques:

These mechanisms aim to leverage mobile agent security level against tampering attacks. Using these techniques will make it very difficult to illegally access or modify the code. Prevention techniques varies according to the goal of prevention which includes: preventing the entire agent or part of it, preventing attacks permanently or temporary, and trusting some functionalities or no trust is assumed [ 24].

Some of these techniques relies on trusted environments equipped with tamper proof hardware computing base at each hosting platform [25]. However, the extension to a tamper proof hardware at each hosting platform limits the ability of a mobile agent to migrate at its will. [26] Used the concept of encrypted functions to prevent code tampering. Their approach encrypts both code and data including state information in a way that enables direct computation on encrypted data without decryption.

In [17], the authors have proposed The use of homomorphic encryption on a part of the data to be executed on the hosts without decrypting it. The main idea is to use a central host as a server which is responsible of encrypting and decrypting operations. Hasegawa et. al. proposed two secure mobile agent protocols with emphasis on efficient oblivious transfer suitable for secure function evaluation in un-trusted environments [27]. However, their model assumes that an agent can travel to one hop and then return back which limits the mobility property of agents.

## 5. Conclusions and Future Work

In this paper, we provide a deep overview of the most recent threats facing the designers of agent platforms and the developers of agent-based applications. The paper identifies security objectives, threats, and countermeasures for these threats. Different security techniques for mobile agents were listed either for tamper detection or prevention. These techniques either inform the occurrence of attack or try to prevent it. Prevention is not always guaranteed and some extra effort should be done to enhance these techniques.

Mobile agents has drawn much attention as a fundamental technology in next generation computing. However, lots of security issues should be well addressed including: on-running agent geo-localization, inter- mobile agent collaboration, real time attack detection, and mutual authentication between host ad mobile agent.

**Belal Amro** is an assistant professor in the Department of Computer Science at Hebron University - Palestine, where he has been working since 2003. Currently, he is conducting research in network security, wireless security, privacy preserving data mining techniques. From 2003 to 2004, he was a research assistant at Hebron University. From 2005 to 2007, he was an instructor in the Computer Science Department at Hebron University after having his MSc. degree of complexity and its interdisciplinary applications form Pavia- Italy. During 2008-2011 he received an ERASMUS PhD grant in Sabanci University-Turkey. From 2011-2012 he worked as research assistant at Sabanci University. In 2012, Belal received a PhD in Computer Science and Engineering From Sabanci University- Istanbul,, turkey.